\documentclass{article}
\usepackage[utf8]{inputenc}
\usepackage[T1]{fontenc}
\usepackage{fvextra}   

\DefineVerbatimEnvironment{PromptBox}{Verbatim}{
  breaklines=true,
  breakanywhere=true,
  fontsize=\footnotesize,
  baselinestretch=1,
  xleftmargin=0.5em,
  frame=single,
  framerule=0.3pt,
  rulecolor=\color{black},
  numbers=left,
  numbersep=4pt
}

    \PassOptionsToPackage{numbers, compress}{natbib}
 \usepackage[preprint]{neurips_2025}

\usepackage{graphicx}
\usepackage{tabularx}
\usepackage[utf8]{inputenc} 
\usepackage[T1]{fontenc}    
\usepackage{hyperref}       
\usepackage{url}            
\usepackage{booktabs}       
\usepackage{amsfonts}       
\usepackage{nicefrac}       
\usepackage{microtype}      
\usepackage{xcolor}         

\providecommand{\cab}{CryptoAnalystBench}

\title{CryptoAnalystBench: Failures in Multi-Tool Long-Form LLM Analysis}

\author{
\textbf{Anushri Eswaran}$^{1,2}$ \\
\textbf{Oleg Golev}$^{1}$ \\
\textbf{Darshan Tank}$^{1}$ \\
\textbf{Sidhant Rahi}$^{1}$ \\
\textbf{Himanshu Tyagi}$^{1}$ \\
\\
$^{1}$ Sentient Labs \\
$^{2}$ University of California, San Diego \\
\texttt{\{oleg, darshan, sidhant, himanshu\}@sentient.xyz} \\
\texttt{aeswaran@ucsd.edu}
}

\begin{document}

\maketitle

\begin{abstract}
Modern analyst agents must reason over complex, high-token inputs—dozens of retrieved documents, tool outputs, and time-sensitive data. While prior work has produced tool-calling benchmarks and examined factuality in knowledge-augmented systems, relatively little work studies their intersection: settings where LLMs must integrate large volumes of dynamic, structured and unstructured multi-tool outputs. We investigate LLM failure modes in this regime using crypto as a representative high-data-density domain. We introduce: (1) CryptoAnalystBench, an analyst-aligned benchmark of 198 production crypto and DeFi queries spanning 11 categories; (2) an agentic harness equipped with relevant crypto/DeFi tools to generate responses across multiple frontier LLMs; and (3) an evaluation pipeline with citation verification and an LLM-as-a-judge rubric spanning four user-defined success dimensions: relevance, temporal relevance, depth, and data consistency. Using human annotation, we develop a taxonomy of seven higher-order error types that are not reliably captured by the factuality checks nor LLM-based quality scoring. We find that these failures persist even in state-of-the-art systems and can compromise high-stakes decisions. Based on the taxonomy, we refine the judge rubric to better capture these errors. While the judge does not align with human annotators on precise scoring across rubric iterations, it reliably identifies critical failure modes, enabling LLM judges to serve as a scalable feedback mechanism for developers and researchers studying analyst-style agents. We release CryptoAnalystBench\footnote{\url{https://github.com/sentient-agi/CryptoAnalystBench}.} with annotated queries, our evaluation pipeline, judge rubrics, and the error taxonomy. Finally, we outline possible error mitigation methods and open challenges in evaluating long-form, multi-tool-augmented systems.
\end{abstract}

\section{Introduction}
Over the past few years, the deployment of large language models (LLMs) as analyst agents has rapidly transitioned from prototypes to production systems that inform high-stakes decisions across finance, law, medicine, and enterprise intelligence. These modern agents routinely process complex workflows that involve dozens of tool calls across heterogeneous sources, including market data APIs, document retrieval systems, and database queries, often aggregating and analyzing hundreds of thousands of tokens of evidence. In finance alone, LLM-powered analyst systems now assist with portfolio management, risk assessment, regulatory compliance, and investment research at major institutions \cite{esma_llm_finance}.

Despite their proliferation in production environments, our understanding of how these agentic systems fail under realistic conditions remains surprisingly incomplete. In general, contemporary analyst agents can be defined by three central challenges:
\begin{enumerate}
    \item They must correctly orchestrate many tools, such as API calls, web searches, document retrievals, and code execution, to gather the necessary evidence. A single analyst query may trigger 10--20 tool invocations, each returning structured or unstructured data.
    \item They must process these high-token-density inputs accurately, synthesizing insights from the dozens of aforementioned tool calls, far exceeding the scale of traditional question-answering benchmarks.
    \item They reason over volatile, time-sensitive data where numeric values and market conditions change quickly within days or hours, demanding temporal awareness, and careful reconciliation of sources based on current relevance.
\end{enumerate}

In this paper, we explore the downstream errors that arise from LLMs having to deal with these challenges, sharing the first perspective of failure modes for these analyst-like agents.

These challenges converge most acutely in domains such as cryptocurrency and decentralized finance (DeFi), where decisions carry significant financial risk and operate on rapidly changing data. Consider a representative query: \textit{"How should I adjust my ETH and SOL portfolio for the next quarter?"} Answering this question requires fetching current and historical price data for Ethereum and Solana, querying blockchain metrics for network activity and validator performance, retrieving recent and future ecosystem developments, performing comparative analysis of market trends and technical indicators, and synthesizing findings into actionable portfolio recommendations---all while maintaining fidelity and correct source attribution.

Any failures of LLMs in synthesizing these data can lead to a poorly-grounded investment recommendation and poor human financial decisions. For example: a hallucinated or incorrectly identified total value locked (TVL) figure may lead to misestimated protocol risk; an incorrectly interpreted yield forecast could result in flawed investment allocation across multiple opportunities; a missed protocol update or event might cause failure to anticipate additional risk. As use cases of such analyst systems are considered by hedge funds, proprietary trading desks, and portfolio managers, understanding their failure modes is critical.

Thus the cryptocurrency domain is particularly well-suited for studying analyst agent reliability. It exhibits extreme data density (dozens of relevant protocols, tokens, and metrics per query), high temporal volatility (market conditions and protocol states evolving hourly), diverse data modalities (price feeds, smart contract code, governance documents, and on-chain analytics), and significant decision stakes (investment and risk management implications). Moreover, crypto analysis requires the full spectrum of analyst capabilities: quantitative reasoning, multi-source synthesis, technical evaluation, and strategic recommendation. These properties make cryptocurrency an ideal testbed for examining failure modes that generalize to other high-stakes, data-intensive domains.

To enable the first systematic investigation of multi-tool analyst agent behaviors under realistic data conditions, we introduce the CryptoAnalystBench benchmark and analyze failure modes that emerge when LLMs conduct complex reasoning with multi-tool orchestration over volatile crypto data.

\section{Related Work}

While evaluation standards for LLM agents is a research area in and of itself \cite{Mohammadi_2025}, relevant literature to this paper can be generally divided into the following four topics: knowledge-augmented LLMs, tool-augmented LLMs, use of LLM-as-a-judge, and the current status of crypto-specific benchmarks. In this section, we briefly survey each of these topics.

\subsection{Knowledge-Augmented LLMs}

Retrieval-augmented generation (RAG) and web-enabled assistants enable inclusion of non-training data into LLM responses. A large body of research has been dedicated to evaluating LLM capabilities' in making grounded claims, avoiding hallucinations, and accurately emitting citations \cite{gao2024retrievalaugmentedgenerationlargelanguage}. For long-form, evidence-grounded systems, rigorous claim verification can be decomposed into three coupled requirements: (i) \emph{claim-to-citation coverage}, i.e., the fraction of claims that are explicitly linked to a citation; (ii) \emph{attribution accuracy}, i.e., whether each cited claim is actually supported by the cited evidence; and (iii) \emph{source correctness}, i.e., whether the cited source is the \emph{correct} source to rely on among multiple sources that may be contradictory.

\subsubsection{Citations \& Attribution}

Attribution is a well-studied property of natural language systems that describes whether every claim within an LLM-generated response is supported by its cited evidence. Early research work in this area introduced new benchmarks and, via human annotation, exposed clear limitations of LLMs in ensuring attribution accuracy in long-form generated texts \cite{liu2023evaluatingverifiabilitygenerativesearch, kamalloo2023hagridhumanllmcollaborativedataset, malaviya2024expertqaexpertcuratedquestionsattributed}.

Automating attribution verification naturally followed, leading to AttributionBench \cite{li2024attributionbenchhardautomaticattribution}, ALCE \cite{gao2023enablinglargelanguagemodels}, and Google's AutoAIS \cite{bohnet2023attributedquestionansweringevaluation}. ALCE in particular is one of the first major papers on the topic and establishes the initial findings of attribution deficiencies in AI texts via automated evaluation \cite{gao2023enablinglargelanguagemodels}. Two years later, SourceCheckup evaluates RAG with medical documents and finds a similar story: the majority (50--90\%) of citations in long-form responses are not fully supported by the cited sources, as verified by human annotators \cite{wu2024llmsciterelevantmedical}. More recently, CiteEval-Auto leverages stronger models and improves verification procedures, reporting that RAG systems can now produce correct context attribution and citations in majority of cases as judged by GPT-4o and human annotators, while still observing that longer generations tend to reduce attribution completeness \cite{xu2025citeevalprincipledrivencitationevaluation}.

Based on this line of work, for this paper we implement an industry-standard claim verification pipeline using a powerful LLM judge model to extract and categorize claims, and verify each claim against adjacent citations when provided.

\subsubsection{Conflict Resolution} When dealing with many tool outputs, it is imperative to assess not only the accuracy of claim attribution but also whether the claim is based on the correct data. This relies on the ability of an LLM to reconcile conflicting information, whether due to quality issues or temporal irrelevance of one or more of the sources. \textit{DRAGged into Conflicts} was one of the first papers to systematize this issue. It presented a taxonomy of conflict types in RAG-augmented LLMs, specified expected model behaviors, and showed how reasoning prompting methods can improve conflict resolution performance \cite{cattan2025draggedconflictsdetectingaddressing}. WikiContradict similarly benchmarks RAG-augmented LLMs on their ability to resolve real-world knowledge conflicts, this time using articles from Wikipedia \cite{hou2024wikicontradictbenchmarkevaluatingllms}.

In this paper, we include source reconciliation as an important category in our error taxonomy for knowledge- and tool-augmented systems. No rigorous work has described this error type in such systems so far.

\subsection{Tool-Augmented LLMs}

Tool-augmented LLMs extend generation with arbitrary external data (API calls, code execution, web search, and more) allowing models to retrieve up-to-date information and operate over structured environments. Early instantiations of such powerful systems included ReAct \cite{yao2023reactsynergizingreasoningacting} and Toolformer \cite{schick2023toolformerlanguagemodelsteach} which decide when to act, what to call, and how to incorporate tool outputs into reasoning or the final answer. This concept of "agent architectures" introduced new requirements for agents to not only produce concrete answers but also to select tools accurately in a desired order and parse their outputs.

Many benchmarks, usually ground-truth-based, have been produced to evaluate LLM capabilities for reliable tool-calling: ToolQA \cite{zhuang2023toolqadatasetllmquestion}, API-Bank \cite{li2023apibankcomprehensivebenchmarktoolaugmented}, RestBench \cite{song2023restgptconnectinglargelanguage}, ToolLLM \cite{qin2023toolllmfacilitatinglargelanguage}, and many others, including subsets within specific industries like finance and crypto. Tool-augmented LLM evaluation also extended to evaluating more generic multi-modal systems, notably GAIA and BrowseComp which stress web- and browser-enabled assistants on real-world information-seeking tasks that require persistence and multi-step navigation \cite{wei2025browsecomp, mialon2023gaia}.

Despite rapid progress, most tool-agent benchmarks still emphasize discrete success criteria and often assume short verifiable traces or final outputs. None provide good visibility into whether agents can reliably \emph{synthesize} large volumes of heterogeneous tool inputs into long-form analytical responses with consistent quantitative claims. This motivates evaluation protocols that go beyond tool-use success to explicitly measure downstream answer quality combined with proper claim attribution.

\subsection{LLM-as-a-Judge}

The majority of latest answer quality evaluations that have not come from tool-calling benchmarks came from LLM-as-a-judge based benchmarks. LLM judges offer a practical alternative to human evaluation for open-ended tasks where reference answers are unavailable or underspecified, by prompting a strong model to provide either pairwise preferences or rubric-based scores. 

A study by Zheng et al. systematically examined this paradigm, showing that strong judges can approximate human preferences (as collected via Chatbot Arena) while also surfacing key failure modes such as position, verbosity, and self-enhancement biases, and proposing mitigations such as balanced-order comparisons and calibration \cite{zheng2023judgingllmasajudgemtbenchchatbot}. G-Eval further introduced a form-filling paradigm in which an LLM uses CoT with explicit evaluation steps to produce discrete rubric scores, improving correlation with human judgments for summarization and dialogue generation over baselines \cite{liu2023geval}. In the aforementioned knowledge-augmented LLM literature, similar findings of good human-LLM alignment are noted \cite{li2024attributionbenchhardautomaticattribution, gao2023enablinglargelanguagemodels, bohnet2023attributedquestionansweringevaluation, wu2024llmsciterelevantmedical, xu2025citeevalprincipledrivencitationevaluation}.

As LLM-as-a-judge has been adopted for fast iteration in model development, subsequent work has emphasized scalability and benchmark construction: BenchBuilder for instance can automatically curate challenging prompts from crowdsourced data and use an LLM judge for fully automated evaluation, reporting strong alignment with human preference rankings while enabling frequent benchmark refreshes \cite{li2024arenahard_benchbuilder}. In addition, several works train evaluator models directly, aiming to make judging cheaper, reproducible, and controllable: Prometheus trains an open evaluator that can score long-form responses using user-provided rubrics \cite{kim2024prometheusinducingfinegrainedevaluation}, and JudgeLM proposes fine-tuning smaller LLMs as cheaper scalable judges to approximate stronger models \cite{zhu2025judgelmfinetunedlargelanguage}.

However, recent work has also demonstrated that LLM judges exhibit systematic biases that can materially affect rankings and absolute scores. Answer length bias is a particularly persistent confounder in preference-based evaluation; Length-Controlled AlpacaEval proposes a simple regression-based adjustment to debias LLM preferences with respect to verbosity \cite{dubois2025lengthcontrolledalpacaevalsimpleway}. Beyond length, multiple studies quantify judge-specific biases including self-preference \cite{wataoka2025selfpreferencebiasllmasajudge}, biases induced by the rubric prompt itself \cite{li2026evaluatingscoringbiasllmasajudge}, and broader families of evaluation biases summarized in automated bias quantification frameworks \cite{ye2024justiceprejudicequantifyingbiases}. Position bias remains a recurrent issue even for strong judges, motivating systematic measurement protocols based on repeated trials and swapped ordering \cite{shi2025judgingjudgessystematicstudy}.

These works indicate that we must treat LLM judges as an \emph{imperfect but scalable} signal, useful for quick feedback or binary identification of issues \cite{zheng2023judgingllmasajudgemtbenchchatbot, shi2025judgingjudgessystematicstudy}. 

\subsection{Crypto and Web3 Benchmarks}

A growing set of benchmarks targets the Web3 and cryptocurrency domain, motivated by its technical complexity, adversarial information landscape, and rapidly shifting market state. However, majority of existing work largely emphasizes questions with simple answers and an objective ground-truth.

CAIA is one of the first data retrieval and manipulation benchmarks which tests LLM systems via ground-truth based evaluation on Web3 reasoning and tool-calling \cite{dai2025caia}. However, its query set is focused on a specific historical market snapshot in 2025, making it less relevant post-publication. On the other hand, CryptoBench introduces a dynamic, expert-curated benchmark with monthly updates (50 questions per month) designed to mirror professional crypto analyst requirements \cite{guo2025cryptobench}. While the queries themselves still focus on short answer generation and tool-calling, a distinguishing feature is its four-quadrant categorization of queries along two axes---retrieval vs.\ prediction and simple vs.\ complex. DMind is one of the more comprehensive benchmarks in crypto with broad Web3 coverage of multi-choice and domain tasks (e.g., contract debugging, DeFi mechanisms, on-chain numeric reasoning) to test both grounded and ungrounded competence across subfields \cite{huang2025dmind}.

Benchmarks like these demand specific APIs like Etherscan, Dune Analytics, and Chainalysis to get the very specific timestamp-based data for short answers. Based on feedback from crypto traders, we realized that the majority of crypto users demand much more open-ended scope from their AI assistants. Benchmarks like CAIA, DMind, and CryptoBench can help us check a crypto agent's tool-calling capability to gather historical data, but not long-form answer quality.

Related trading-oriented work includes InvestorBench, which evaluates agents across diverse financial decision-making contexts including cryptocurrencies \cite{li2024investorbenchbenchmarkfinancialdecisionmaking}, and CryptoTrade, which benchmarks LLM-based trading strategies combining on-chain and off-chain signals \cite{li2024cryptotrade}. These benchmarks assume sharp action outputs from LLMs rather than analyses and are thus not helpful for evaluating long-form conversational agent outputs as-is.

\section{Contributions}
We make the following contributions:

\textbf{(1) CryptoAnalystBench: A Production-Aligned Benchmark.} We introduce a benchmark of 198 carefully curated
queries spanning 11 categories of cryptocurrency and DeFi
analysis, including protocol comparison, risk assessment,
yield optimization, governance evaluation, security analysis, and market dynamics. Unlike synthetic benchmarks, our
queries are derived from real analyst workflows and designed
with time-robust framing: questions remain meaningful as
markets evolve, avoiding dependence on specific price levels or transient events while still requiring reasoning over
current data. We also introduce an LLM-as-a-judge
rubric grounded by expert crypto analyst input.

\textbf{(2) Agentic Evaluation Harness with Multi-Tool Infrastructure.} We develop a comprehensive evaluation infrastructure equipped with production-grade tools, including
market data APIs (CoinGecko, DefiLlama), web search, document retrieval systems, blockchain query interfaces, and
code execution environments. This harness mirrors the capabilities available to production analyst agents and enables
controlled experimentation across frontier LLMs. We collect
long-form responses reflecting real-world analyst outputs,
yielding a dataset suitable for studying extended reasoning
failure patterns.

\textbf{(3) Multi-Level Evaluation Methodology.} We propose a layered assessment framework combining two complementary
approaches:
\begin{itemize}
\item Automated citation verification, which extracts claims, maps
them to tool outputs, and flags unsupported assertions in
multi-source, structured-data settings;
\item LLM-as-a-judge evaluation, employing rubric-based scoring across four dimensions: depth, query relevance, temporal relevance, and data consistency;
\end{itemize}
This multi-level approach enables empirical analysis of both
the strengths and blind spots of automated evaluation techniques.

\textbf{(4) Failure Mode Taxonomy and Comparative Analysis.}
Through iterative analysis of model outputs, automated signals, and human expert annotations, we develop a hierarchical failure taxonomy distinguishing surface-level errors (e.g., factual inaccuracies, missing citations) from higher-order structural failures unique to multi-tool workflows.

Taken together, our benchmark, evaluation pipeline, and failure taxonomy provide a practical methodology for diagnosing reliability bottlenecks in long-form, tool-augmented analyst systems operating over dynamic, high-volume data.

\section{The CryptoAnalystBench Benchmark}
Partially inspired by other benchmarks \cite{guo2025cryptobench,dai2025caia,bigeard2025financeagentbench}, CryptoAnalystBench follows five principles:

\begin{enumerate}
\item Production realism: queries are drawn from real user traffic
to capture actual demand distribution and language.
\item Analyst-aligned evaluation: the harness scores what users
value in a long-form copilot response: relevance, depth, temporal accuracy, and internal consistency.
\item Representativeness on topic frequency: we clustered
production queries and used cluster sizes to determine the
benchmark's query type distribution.
\item Time-robust question framing: questions are designed
to remain meaningful over time (answers may vary, but the
question remains valid).
\item Scalable evaluation: LLM-as-a-judge enables large-scale comparison of open-ended outputs, with bias controls
\cite{li2024llmsasjudges}.
\end{enumerate}
\begin{table}[!h]
\centering
\caption{\textbf{Benchmark construction pipeline.}}
\label{tab:construction}
\small
\setlength{\tabcolsep}{4pt}
\renewcommand{\arraystretch}{1.2}
\begin{tabularx}{\columnwidth}{@{}l l X@{}}
\toprule
\textbf{Stage} & \textbf{Output} & \textbf{Description} \\
\midrule
1.\ Filtering &
Raw pool &
Remove overly long instructions and under-informative short queries (2--3 words)
\\
2.\ Clustering &
$\sim$2{,}000 queries &
Semantic grouping into 11 analyst-aligned categories
\\
3.\ Diversity selection &
$\sim$650 queries &
LLM-assisted removal of near-duplicates differing only in entities (e.g., token tickers)
\\
4.\ Difficulty filtering &
$\sim$350 queries &
Remove basic concept questions answerable without retrieval or tool use
\\
5.\ Expert curation &
198 queries &
Final selection maintaining category proportions and requiring complex reasoning
\\
\bottomrule
\end{tabularx}
\end{table}

\subsection{Benchmark Construction}
CryptoAnalystBench is derived from real production traffic through a five-stage curation pipeline (Table 1). Starting from raw user queries, we
iteratively filtered for diversity, difficulty, and analyst-alignment,
ultimately selecting 198 queries that require web search, tool calling,
and synthesis capabilities rather than simple factual recall. We remove personally identifying information and exclude queries containing sensitive user data. The released benchmark contains only de-identified prompts and does not include user metadata.

\begin{table}[h!]
\centering
\caption{\textbf{Representative queries by category.}}
\label{tab:examples}
\small
\setlength{\tabcolsep}{4pt}
\renewcommand{\arraystretch}{1.2}
\begin{tabularx}{\columnwidth}{@{}lX@{}}
\toprule
\textbf{Category} & \textbf{Example queries} \\
\midrule
\textbf{Project \& Fundamental Research} &
\textit{``Who is backing \$ASTER?''} \newline
\textit{``What is the launch and development roadmap of Monad, and what milestones have been achieved so far?''} \newline
\textit{``How does Solana perform against competitors in key metrics like holder growth or community engagement?''}
\\
\midrule
\textbf{Market Data \& Price Discovery} &
\textit{``What is the all-time high price of Solana, and is there potential for it to reach it again?''} \newline
\textit{``Which projects are trending on DEX volume today?''} \newline
\textit{``Filter tokens with 100M--1B market cap and $>$10\% 24h gain''}
\\
\midrule
\textbf{On-Chain Analytics \& Flows} &
\textit{``Are ETH whales accumulating today?''} \newline
\textit{``What's happening with DeFi liquidations this week?''} \newline
\textit{``Are there staking, reflections, or airdrop opportunities for holders of Monad?''}
\\
\midrule
\textbf{Macro \& Narrative Context} &
\textit{``Identify extreme fear moments in the crypto market over the last 3 days''} \newline
\textit{``How might Polygon's cross-chain integration affect the DAO market?''} \newline
\textit{``Why is \$aster going up?''}
\\
\bottomrule
\end{tabularx}
\end{table}

\subsection{Question Taxonomy}
We group queries into 11 categories that capture the vast majority of
investment-relevant intents observed in our production traffic (e.g.,
market data and price discovery, on-chain analytics and flows,
project and fundamental research). The distribution across categories is shown in Figure 1.

\begin{figure}[!h]
  \centering
  \includegraphics[width=1.0\linewidth]{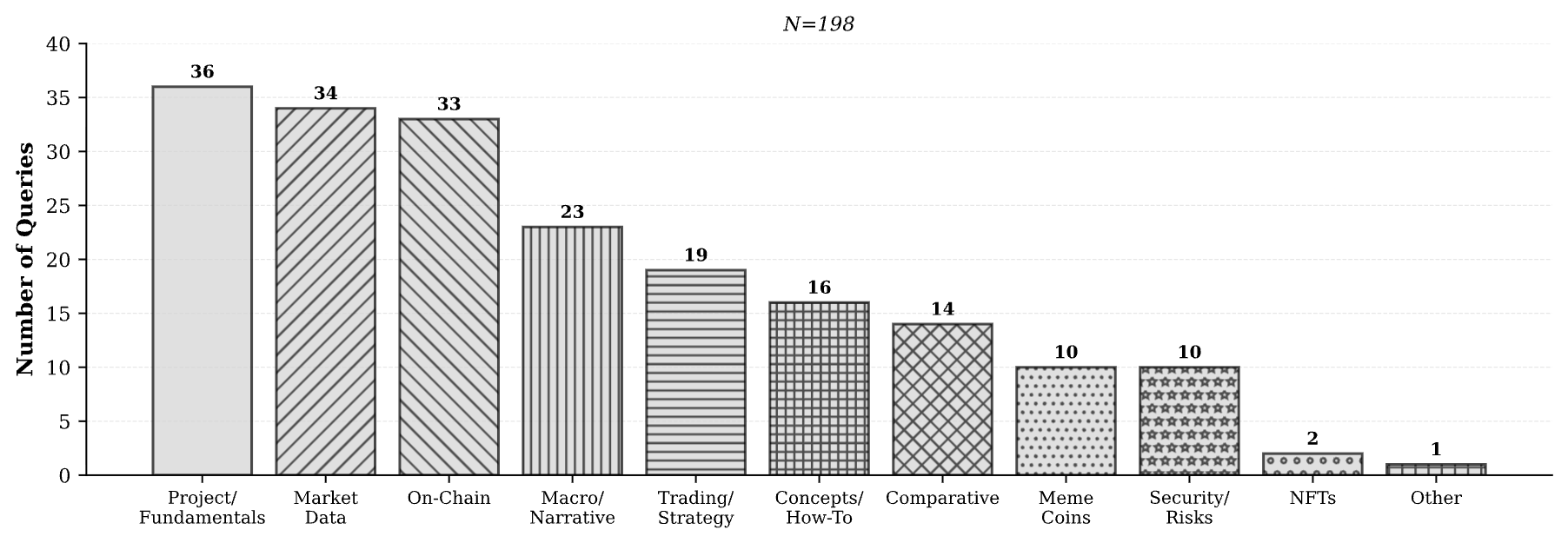}
  \caption{\textbf{Category distribution of \cab queries.}}
  \label{fig:category_dist}
\end{figure}

To illustrate the practical scope of CryptoAnalystBench, Table 2 presents representative queries from the four largest categories. These 
examples reflect production analyst workflows spanning real-time
market monitoring, on-chain interpretation, project due diligence,
and macro narrative analysis.

\begin{figure}[!h]
  \centering
  \includegraphics[width=0.7\linewidth]{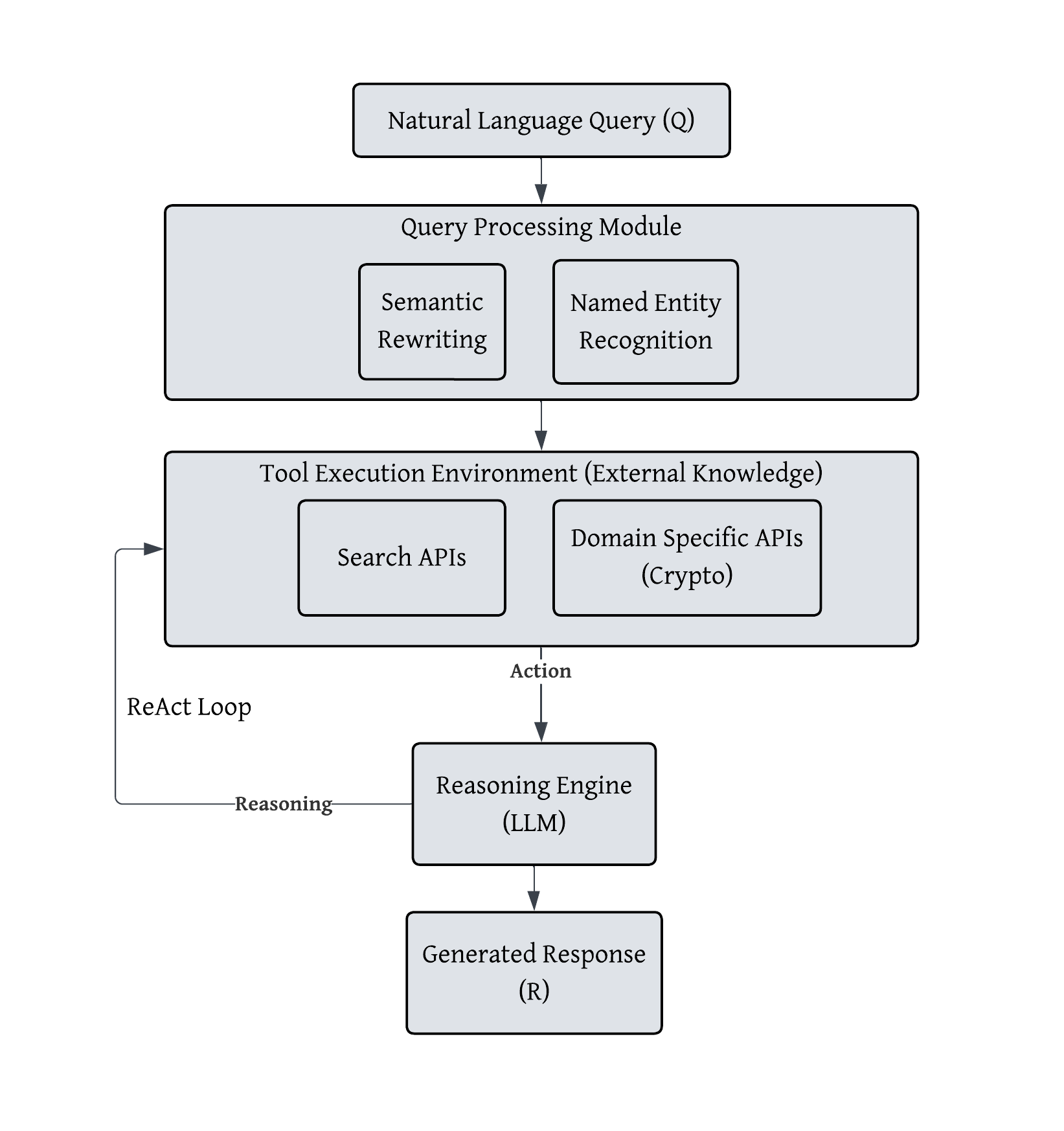}
    \caption{\textbf{Agentic Harness} architecture for evaluating CryptoAnalystBench, illustrating query processing, tool execution, and ReAct-style reasoning.}
  \label{fig:architecture_diag}
\end{figure}

\section{Agentic Harness}

To evaluate LLM performance on CryptoAnalystBench, we design an agentic evaluation harness that closely mirrors production-grade cryptocurrency analyst systems. As shown in Figure~\ref{fig:architecture_diag}, a natural language query is first processed by a query understanding module that performs semantic rewriting and named entity recognition. The processed query is then passed to a tool execution environment, where the agent can invoke external knowledge sources, including general-purpose search APIs and domain-specific cryptocurrency APIs. Guided by the reasoning LLM, tool outputs are iteratively incorporated through a ReAct-style loop, enabling structured data gathering and multi-step reasoning prior to producing the final generated response.

\section{Evaluation}
\label{sec:llm_judge}

We use LLM-as-a-judge with structured JSON outputs to score responses across four dimensions
Appendix~\ref{app:llmprompt} described in Table \ref{tab:eval_criteria}.

\begin{table}[!h]
\centering
\caption{\textbf{Evaluation dimensions (1--10 scale).}}
\label{tab:eval_criteria}
\small
\setlength{\tabcolsep}{6pt}
\renewcommand{\arraystretch}{1.2}
\begin{tabular}{@{}lp{0.65\columnwidth}@{}}
\toprule
\textbf{Dimension} & \textbf{Description} \\
\midrule
Temporal relevance (TR) &
Freshness of data relative to the evaluation date \\
Data consistency (DC) &
Internal coherence and absence of contradictions \\
Depth (D) &
Comprehensiveness, technical detail, and multi-axis coverage \\
Relevance (R) &
Directness in addressing user intent and decision context \\
\bottomrule
\end{tabular}
\end{table}

\begin{table}[!h]
\centering
\caption{\textbf{Model performance across evaluation metrics (1--10 scale).}}
\label{tab:model_performance}
\small
\setlength{\tabcolsep}{6pt}
\renewcommand{\arraystretch}{1.2}
\begin{tabular}{@{}lcccc@{}}
\toprule
\textbf{Model} & \textbf{Temporal Relevance} & \textbf{Data Consistency} & \textbf{Depth} & \textbf{Relevance} \\
\midrule
GPT OSS 120B & 8.27 & 8.77 & 7.25 & 9.17 \\
GLM 4.7      & 8.46 & 8.99 & 7.85 & 9.27 \\
Kimi k 2.5   & 8.78 & 9.03 & 8.35 & 9.58 \\
Qwen 235B    & 8.49 & 8.84 & 8.01 & 9.36 \\
GPT-5.2      & 8.77 & 9.21 & 8.15 & 9.57 \\
\bottomrule
\end{tabular}
\end{table}

The results in Table~\ref{tab:model_performance} highlight distinct optimization profiles across models. Temporal relevance and query relevance appear largely saturated across all systems, whereas depth and data consistency emerge as the primary axes of differentiation. Notably, GPT-5.2 emphasizes consistency, while Kimi K2.5 achieves greater depth without sacrificing relevance, making it a strong generalist choice for agent-based workloads. 

To validate the reliability of the LLM-as-a-judge evaluation, we conducted a human assessment using manual annotators with domain expertise in crypto markets. The annotators evaluated the outputs of one selected model and their judgments were compared against the LLM-based scores. All annotators were provided with detailed descriptions of the four evaluation rubrics prior to annotation.

For each query, we run the agentic harness to produce a single long-form response per model under a fixed prompt and tool set. We report mean rubric scores across all benchmark queries. For human evaluation, expert annotators score a subset of model outputs using the same rubric definitions; we report agreement using quadratic-weighted Cohen's Kappa (Table ~\ref{tab:kappa}).
Importantly, this human evaluation is not intended to establish a gold-standard calibration for scalar scores, but rather to contextualize score-level disagreement in open-ended analysis.

\begin{table}[h]
\centering
\caption{\textbf{Cohen’s Kappa (quadratic weights)} measuring agreement between LLM-as-a-judge scores and human expert annotations across evaluation rubrics.}
\begin{tabular}{lc}
\hline
\textbf{Dimension} & \textbf{$\kappa_w$} \\
\hline
temporal relevance & 0.3856 \\
data consistency   & 0.3493  \\
depth               & 0.4178 \\
relevance           & 0.3589 \\
\hline
\end{tabular}
\label{tab:kappa}

\end{table}

\begin{table}[h]
\centering
\caption{\textbf{Seven-category taxonomy of qualitative failure modes identified.}}
\begin{tabular}{p{0.32\linewidth} p{0.6\linewidth}}
\hline
\textbf{Category} & \textbf{Description} \\
\hline
Staleness / Missing Time Bounds & Information is outdated or lacks explicit temporal grounding, leading to unclear validity windows. \\
Inconsistent Claims & Internal contradictions or mutually incompatible assertions within the same response. \\
Source Reconciliation Failure & Inability to resolve or acknowledge conflicts between multiple cited or implied sources. \\
Shallow Synthesis & Surface-level aggregation of facts without meaningful integration or reasoning. \\
Missing Risk or Mechanism Context & Omission of underlying mechanisms, assumptions, or risk factors necessary for proper interpretation. \\
Overconfident Prediction & Strong claims made without sufficient evidence, uncertainty calibration, or stated assumptions. \\
Partial or Misframed Answer & Response addresses only part of the query or interprets the question incorrectly. \\
\hline
\end{tabular}
\label{tab:error_taxonomy}

\end{table}

Despite reasonable alignment, the LLM-as-a-judge scores exhibit only fair-to-moderate agreement with human experts under weighted Cohen’s Kappa, reflecting the inherently subjective nature of open-ended answers and differences in calibration between human and model judgments. This motivates shifting the role of human annotation from supervising scores to defining and validating recurring error types. To better characterize these discrepancies, we introduce a seven-layered error taxonomy that captures recurring failure modes observed during human annotation (Table~\ref{tab:error_taxonomy}). These categories isolate distinct qualitative shortcomings—ranging from temporal ambiguity to overconfident extrapolation—that are not fully captured by scalar rubric scores alone. After manually annotating instances of each failure type, we trained and deployed an LLM-based classifier to automatically identify these categories. Validation against human annotations shows that the LLM achieves an identification accuracy of 93.45\%, indicating that while holistic scoring remains subjective, structured error attribution can be reliably automated.

\textbf{Example.} For a query such as ``Which altcoins are breaking all-time highs this week?'', a model may produce a polished table of assets and prices sourced from multiple tools, yet fail to reconcile conflicting ATH values reported by different sources (e.g., listing an ATH of \$0.195 for a token while another tool reports \$0.1812). When such discrepancies are neither acknowledged nor resolved, the response exhibits a \emph{Source Reconciliation Failure}, despite appearing comprehensive and authoritative. Additional examples for each failure category are provided in Appendix~\ref{app:failure_examples}.


\subsection{Citation and Hallucination Verification}
A central challenge in evaluating long-form analytical responses is determining whether individual claims are grounded in retrieved evidence or fabricated by the model. Manually verifying hundreds of factual assertions across dozens of tool invocations is prohibitively expensive, while automated factuality verification remains an open research problem—particularly for multi-source reasoning involving structured data, numerical claims, and temporal dependencies.

To address this challenge, we propose an agentic citation verification pipeline that systematically validates the provenance of claims in model-generated responses. The pipeline consists of three stages. First, claim extraction decomposes long-form responses into atomic factual assertions using an LLM-based segmentation approach. Second, claim typing and knowledge linking classifies extracted claims into exact and derived facts and explicitly links them to external knowledge traces captured during tool usage (e.g., web search results, API outputs, or retrieved documents). Third, source attribution maps each claim to the specific tool outputs that may support it, enabling fine-grained assessment of evidentiary grounding.

From this process, we compute three factuality metrics: \emph{exact} claims, which are directly supported by retrieved tool outputs; \emph{derived} claims, which can be inferred from retrieved evidence via aggregation or simple reasoning; and \emph{fabricated} claims, for which no supporting evidence exists in any retrieved output.

In parallel, we develop a complementary pipeline to evaluate the citation preciseness of model-generated responses. Citation preciseness measures the proportion of extracted factual claims that are accompanied by an explicit source citation, reflecting attribution coverage rather than semantic correctness.

This automated verification process enables us to identify responses that appear fluent and authoritative but lack empirical backing---a particularly insidious failure mode in high-stakes domains where decision-makers may not have the time or expertise to independently verify every assertion. Prompts are in Appendix~\ref{app:citation_verification},~\ref{app:hallucination}.




\section{Results and Analysis}

\begin{figure}[!h]
    \centering
    \includegraphics[width=0.6\textwidth]{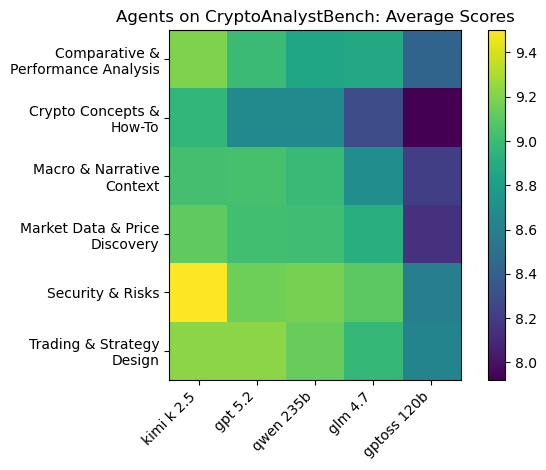}
    \caption{\textbf{Heatmap} of average performance across all four evaluation metrics (temporal relevance, data consistency, depth, and relevance) for each model and domain tag. Higher values indicate stronger overall agent performance.}
    \label{fig:avg_metric_heatmap}
\end{figure}

Figure~\ref{fig:avg_metric_heatmap} illustrates aggregated performance across all evaluation metrics, highlighting clear cross-domain specialization patterns. Notably, Kimi K2.5 maintains uniformly strong performance across domains, whereas models like GPT-5.2 demonstrate competitive but more task-dependent behavior.

Table~\ref{tab:citation_hallucination} shows that all models predominantly rely on tool-grounded or correctly derived information, with fabricated claims remaining below 6\% across the board. Qwen-235B and GLM-4.7 exhibit the low hallucination rates, while GPT-5.2 and Kimi K2.5 balance strong grounding with higher reliance on model-derived knowledge.

\begin{table}[!h]
\centering
\caption{\textbf{Hallucination attribution breakdown across models.}}
\label{tab:citation_hallucination}
\small
\setlength{\tabcolsep}{6pt}
\renewcommand{\arraystretch}{1.2}
\begin{tabular}{@{}lccc@{}}
\toprule
\textbf{Model} & \textbf{Exact \%} & \textbf{Derived \%} & \textbf{Fabricated \%} \\
\midrule
GPT OSS 120B & 61.72 & 32.90 & 5.49 \\
Qwen 235B    & 74.59 & 23.36 & 1.76 \\
Kimi K2.5    & 66.98 & 31.44 & 1.51 \\
GPT-5.2      & 66.42 & 32.02 & 1.55 \\
GLM 4.7      & 70.18 & 28.63 & 1.22 \\
\bottomrule
\end{tabular}
\end{table}

Table~\ref{tab:citation_preciseness} reports citation preciseness across models, measuring whether each factual claim in a response is explicitly accompanied by a source citation. Overall, performance is strong, with all evaluated models achieving preciseness above 85\%, indicating that the majority of asserted facts are properly attributed. GLM-4.7 and Kimi achieve the highest citation preciseness, suggesting more consistent attribution of factual claims, while GPT-5.2 closely follows with comparable performance. In contrast, GPT-OSS-120B and Qwen-235B show slightly lower preciseness, reflecting a higher incidence of uncited factual assertions despite otherwise grounded responses.

\begin{table}[t]
\centering
\caption{\textbf{Citation preciseness across evaluated models.}}
\label{tab:citation_preciseness}
\small
\setlength{\tabcolsep}{6pt}
\renewcommand{\arraystretch}{1.2}
\begin{tabular}{@{}lc@{}}
\toprule
\textbf{Model} & \textbf{Citation Preciseness (\%)} \\
\midrule
GPT OSS 120B & 89.88 \\
Qwen 235B    & 86.95 \\
Kimi K2.5    & 95.69 \\
GPT-5.2      & 95.50 \\
GLM 4.7      & 97.14 \\
\bottomrule
\end{tabular}
\end{table}

\begin{table}[t]
\centering
\caption{\textbf{Proportion of responses exhibiting each failure category across models.}}
\label{tab:failure_by_model}
\small
\setlength{\tabcolsep}{5pt}
\renewcommand{\arraystretch}{1.2}
\begin{tabular}{@{}lccccc@{}}
\toprule
\textbf{Failure Category} & 
\textbf{GPT OSS 120B} & 
\textbf{GLM 4.7} & 
\textbf{Kimi K2.5} & 
\textbf{Qwen 235B} & 
\textbf{GPT-5.2} \\
\midrule
Staleness / Time     & 0.141 & 0.136 & 0.086 & 0.172 & 0.045 \\
Inconsistent         & 0.025 & 0.025 & 0.010 & 0.035 & 0.020 \\
Source Recon.        & 0.076 & 0.071 & 0.056 & 0.096 & 0.056 \\
Shallow Synth.       & 0.025 & 0.005 & 0.015 & 0.000 & 0.000 \\
Missing Risk         & 0.167 & 0.146 & 0.096 & 0.126 & 0.101 \\
Overconfident        & 0.005 & 0.005 & 0.000 & 0.010 & 0.000 \\
Partial / Misframed  & 0.101 & 0.162 & 0.141 & 0.126 & 0.106 \\
\bottomrule
\end{tabular}
\end{table}

Table~\ref{tab:failure_by_model} shows that while all models exhibit non-trivial rates of contextual and framing errors, the composition of these failures varies substantially by model. GPT-OSS-120B displays a broadly distributed error profile, with elevated rates of temporal staleness and missing risk or mechanism context, suggesting adequate retrieval but weaker global structuring of responses. GLM-4.7 follows a similar pattern but with a notably higher incidence of partial or misframed answers, indicating a tendency toward incomplete synthesis rather than explicit factual error. Kimi-K2.5 exhibits reduced staleness and inconsistency relative to larger models, but still shows a substantial fraction of partial or under-scoped responses, consistent with more conservative or fragmented reasoning strategies. In contrast, Qwen-235B presents the highest rate of temporal staleness and source reconciliation failures, alongside near-zero shallow synthesis, implying aggressive aggregation of retrieved information that increases integration error rather than superficial summarization. GPT-5.2 demonstrates the lowest overall rates of staleness, overconfidence, and shallow synthesis, with residual failures concentrated in missing risk context and partial framing, suggesting a bias toward cautious, well-bounded outputs that occasionally underspecify broader mechanisms. Collectively, these patterns indicate that as models improve factual stability, failure modes shift away from contradictions and hallucinations toward higher-order issues of temporal grounding, explanatory completeness, and framing.

To address these failure modes, we introduce three targeted mitigations: (i) enforcing explicit prioritization of structured API outputs over unstructured search results, (ii) dynamically augmenting prompts with temporal context derived from the query and tool traces, and (iii) selectively activating task-specific sub-prompts to emphasize analytical depth where required.

\begin{figure}[t]
    \centering
    \includegraphics[width=\linewidth]{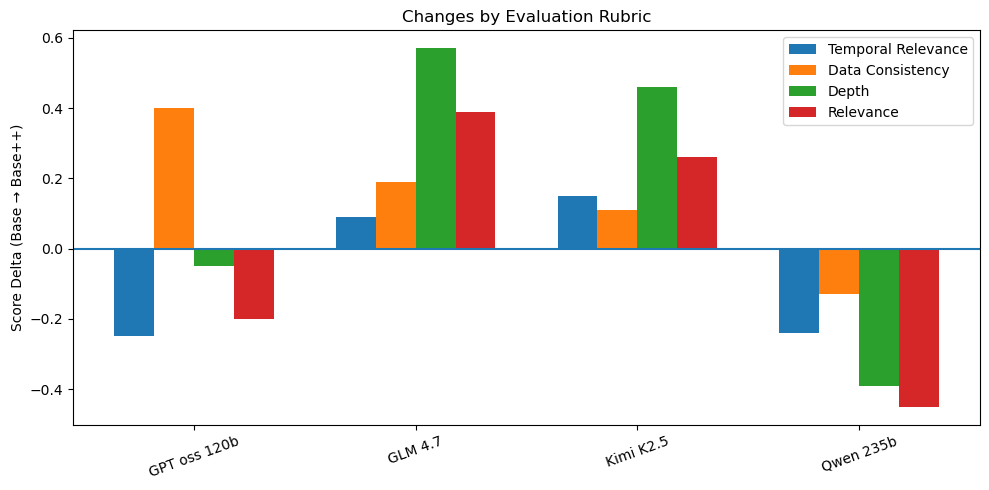}
    \caption{
    Evaluation on Agent Harness with interventions. Positive values indicate improvements while negative values indicate regression
    }
    \label{fig:basepp_rubric_changes}
\end{figure}

Figure~\ref{fig:basepp_rubric_changes} shows the impact of these interventions across evaluation rubrics. We observe consistent gains in analytical depth and relevance for higher-capability models, suggesting that mitigations successfully unlock latent reasoning capacity when the underlying planner is sufficiently robust. However, weaker models exhibit regressions across multiple dimensions, particularly depth and relevance, indicating that additional structural and contextual constraints can overburden limited planners. These results suggest that, while the proposed mitigations are effective, their application should be model-aware rather than uniformly applied. We emphasize that these mitigations are not intended as a general solution, but rather as empirically grounded heuristics that proved effective within our experimental setup.
\section{Future Work}

Several directions emerge from this study. First, improving LLM-as-a-judge reliability remains an open challenge; future work could explore hybrid judging approaches that combine rubric-based scoring with structured failure detection or uncertainty-aware calibration. Second, while our failure taxonomy captures dominant qualitative error patterns, extending it to finer-grained or domain-specific subcategories may further improve diagnostic resolution. Third, adaptive mitigation strategies—where interventions are dynamically selected based on model capability, task complexity, or real-time tool traces—represent a promising avenue for reducing regressions in weaker planners. Although we focus on cryptocurrency as a representative high-data-density domain, applying CryptoAnalystBench-style evaluation to other settings such as legal analysis, enterprise intelligence, or scientific research would help assess the generality of the observed failure modes and guide the development of more robust multi-tool reasoning systems.

\section{Conclusion}

In this work, we present CryptoAnalystBench, a production-aligned benchmark designed to surface failure modes in long-form, multi-tool LLM-based analyst systems operating under realistic data density and temporal volatility. Through a layered evaluation framework combining LLM-as-a-judge scoring, citation verification, and expert-driven failure taxonomy analysis, we show that standard automated metrics alone are insufficient to capture higher-order structural errors. Our findings reveal that even frontier models struggle primarily with contextual framing—such as temporal grounding, risk articulation, and multi-source reconciliation—rather than with isolated factual correctness. Overall, this work highlights the importance of diagnostic, failure-centric evaluation for deploying analyst agents in high-stakes domains and provides both tools and empirical insights to support more reliable system design.

\bibliographystyle{ACM-Reference-Format}
\bibliography{references}

\section{Appendix}
\appendix
\section{Failure Case Examples}
\label{app:failure_examples}

This appendix provides representative examples for each qualitative failure category described in Table~6. Each example illustrates how fluent, citation-rich responses can nonetheless exhibit higher-order structural failures when integrating multi-tool outputs.

\subsection{Staleness / Missing Time Bounds}
\textbf{Query:} Which altcoins are outperforming Bitcoin today?

\textbf{Observed Failure.} The response relied on outdated 24-hour price change data sourced from web search results, while current API data indicated materially different values for Bitcoin and several altcoins.

\subsection{Inconsistent Claims}
\textbf{Query:} What does the head-and-shoulders pattern on Dogecoin's daily chart suggest about increased volatility and potential capital inflow for Ripple?

\textbf{Observed Failure.} The response claimed an inverted head-and-shoulders pattern while citing sources that described a standard bearish formation, resulting in internal contradiction.

\subsection{Source Reconciliation Failure}
\textbf{Query:} Which altcoins are breaking all-time highs this week?

\textbf{Observed Failure.} Conflicting ATH values for the same token across sources were not reconciled or acknowledged.

\subsection{Shallow Synthesis}
\textbf{Query:} Which crypto influencers are pushing new tokens?

\textbf{Observed Failure.} The response listed influencers without analyzing patterns, incentives, or verification mechanisms.

\subsection{Missing Risk or Mechanism Context}
\textbf{Query:} Which crypto influencers are pushing new tokens?

\textbf{Observed Failure.} The response omitted discussion of risks such as undisclosed promotions or incentive misalignment.

\subsection{Overconfident Prediction}
\textbf{Query:} How will Binance's listing of a popular token trading pair draw market attention?

\textbf{Observed Failure.} Precise price forecasts were provided without uncertainty bounds or probabilistic framing.

\subsection{Partial or Misframed Answer}
\textbf{Query:} Which altcoins are breaking all-time highs this week?

\textbf{Observed Failure.} Only a subset of qualifying assets was identified despite broader supporting data.

\section{LLM-as-a-Judge Evaluation Prompt}
Model: Deepseek V3.1
\label{app:llmprompt}

\begin{PromptBox}
EVALUATION DATE AND TIME: {eval_datetime}
QUERY: {query}

CONTEXT THAT WAS AVAILABLE TO THE MODEL
(Evaluate factual accuracy and relevance against this context only; do not use external knowledge.)
---
{trace_context}
---

RESPONSE #{response_num}:
{response}

Please score this response on the four parameters below using a 1--10 scale and provide comprehensive reasoning for each score.

IMPORTANT NOTE ON HISTORICAL QUERIES:
If the query explicitly asks for historical data (e.g., "What was Bitcoin's price in 2017?", "Show me historical trends", "How did X perform last year?"), providing accurate historical information is appropriate and should NOT be penalized on any metric.
For such queries, evaluate whether the historical data provided is accurate and relevant to what was asked.

EVALUATION CRITERIA:

1. TEMPORAL RELEVANCE (1--10):
- Reflects recent data, events, or market conditions
- Up-to-date for crypto/blockchain context
- Avoids outdated or stale information
- Considers current date ({eval_datetime})
- For historical queries: accuracy for the requested time period

2. DATA CONSISTENCY (1--10):
- No contradictions between claims
- Logical alignment of statements and conclusions
- Consistent numbers, dates, and metrics
- No internal logical inconsistencies
- Maintains internal coherence

3. DEPTH (1--10):
- Sufficient technical detail
- Clear explanation of complex concepts
- Organized structure (analysis, data, risks)
- Clear presentation of formulas and metrics
- Covers all relevant aspects of the query

4. RELEVANCE (1--10):
- Directly answers the question
- Practical value for decision-making
- Highlights risks and limitations
- Applicable to real-world crypto scenarios
- Aids informed understanding
- For historical queries: addresses historical intent

REASONING REQUIREMENTS:
For each metric, your reasoning must include:
1. Why the specific score was assigned
2. What is strong about the response and what is missing

OUTPUT FORMAT CONSTRAINT:
Respond with ONLY valid JSON in English.
- Do not include any text before or after the JSON
- Do not use Markdown formatting

REQUIRED JSON SCHEMA:
{
  "temporal_relevance": {
    "score": <1-10>,
    "reasoning": "<detailed explanation>"
  },
  "data_consistency": {
    "score": <1-10>,
    "reasoning": "<detailed explanation>"
  },
  "depth": {
    "score": <1-10>,
    "reasoning": "<detailed explanation>"
  },
  "relevance": {
    "score": <1-10>,
    "reasoning": "<detailed explanation>"
  }
}
\end{PromptBox}

\section{Citation and Hallucination Verification Prompt}
Model: Deepseek V3.1

\label{app:citation_verification}

\subsection*{System Prompt}
\begin{PromptBox}
You are a citation verification expert. Your role is to meticulously evaluate whether citations in LLM responses accurately reference their source materials.

Core Principles:
- Verify EVERY claim against its cited source(s)
- Classify ALL cited claims (no "unclear" categories)
- Be thorough but lenient with paraphrasing and derivations
- Count accurately and ensure numbers add up correctly

CRITICAL: Your reply MUST end with the required JSON block (Part 2).
Without it the evaluation cannot be recorded. Never omit the JSON.
\end{PromptBox}

\subsection*{User Prompt}
\begin{PromptBox}
# Citation Verification Task

Evaluate the accuracy and completeness of citations in the LLM Response by verifying them against the provided Context.

## Context Format Understanding

The Context contains sources in two formats:

1. Detailed Sources:
- Start with "From [source_url]" or "**From [source_url]:**"
- Content follows below the marker
- May use IDs like src_1, src_2, etc.

2. Search Result Snippets (one-liner format):
- Format: [Title]... [Snippet text]. Source: [source_url]
- The snippet text BEFORE "Source:" is verifiable content
- The URL AFTER "Source:" is the correct citation for that snippet
- Claims matching snippets should cite the URL following that snippet

## Evaluation Process

### Step 1: Parse Context
- Identify all sources and their content
- Map source IDs to URLs/content
- Note search result snippets and their associated URLs

### Step 2: Extract ALL Claims
Identify every factual claim in the Response:
- Numbers (prices, percentages, metrics, TVL, volumes)
- Dates and time periods
- Names (people, companies, organizations)
- Technical details and features
- Any quantitative or specific data

For each claim, note:
- The claim content
- Citation status (cited or uncited)
- Citation reference (e.g., src_1, src_2)

### Step 3: Verify Each Cited Claim
For EVERY cited claim:

a) Locate cited source(s): Find the citation reference and corresponding source in Context

b) Verify claim exists: Check if the claim is:
- CORRECTLY CITED (direct): Exact or semantically equivalent match in source
- CORRECTLY CITED (derivable): Can be derived through:
  - Mathematical derivations (e.g., "2x" from "doubled")
  - Unit conversions (e.g., "1.36B" from "1360000000")
  - Approximations (e.g., "~$1B" from "$950M")
  - Format variations (e.g., "19.7%" from "19.683%" rounded)
  - Reasonable interpretations/summaries
  - Synthesis from multiple cited sources
- INCORRECTLY CITED: Cannot be found or derived from cited source(s)

c) For multiple cited sources: Verify claim is supported by combining information from ALL cited sources

Critical: You MUST classify every cited claim into one of the three categories above.

### Step 4: Identify Missing Citations
For each uncited claim:
- Does it need attribution?
- Which source(s) in Context support it?
- Record as missing citation if attribution is needed

### Step 5: Calculate Metrics
Count accurately:
- Total claims identified: All factual claims in Response
- Claims with citations: Claims that have citations
- Correctly cited (direct): Exact matches in cited sources
- Correctly cited (derivable): Derivable from cited sources
- Incorrectly cited: Not found/derivable in cited sources
- Missing citations: Need citations but lack them
- No citation needed: Don't require citations

Verify:
- Claims with citations = Correctly cited (direct) + Correctly cited (derivable) + Incorrectly cited

Citation precision:
- ((Correctly cited direct + Correctly cited derivable) / Claims with citations) x 100%

Citation completeness:
- (Claims with citations / (Total claims - No citation needed)) x 100%

### Step 6: Determine Label
- correct (1.0): ALL claims cited correctly, 100% precision and completeness
- partially correct (0.5): MOST claims cited correctly, precision or completeness > 50% but < 100%
- incorrect (0.0): MANY missing/incorrect citations, precision or completeness <= 50%

## Be Lenient - Do NOT Penalize:
- Paraphrasing or rewording (if core claim is preserved)
- Different formatting of same data
- Reasonable interpretations
- Mathematical derivations from cited sources
- Multiple sources cited together for combined support

## Response Format

### Part 1: Chain of Thought (Text Format)

CHAIN OF THOUGHT:

For each claim:

1. [Claim Name/Type]:
   - Response claims: "[exact claim text]"
   - Has citation: [YES - source(s) / NO]
   - Cited source(s): [source ID(s) or N/A]
   - Source check: "[what source says or NOT FOUND]"
   - Derivation check: "[how it can/cannot be derived if not exact]"
   - Status: [CORRECTLY CITED (direct) / CORRECTLY CITED (derivable) / INCORRECTLY CITED / MISSING CITATION / NO CITATION NEEDED]

[Continue for ALL claims]

### Part 2: Structured Results (JSON Format) — MANDATORY

STRICT JSON OUTPUT RULES:
1. Output the JSON object immediately after your chain of thought. Do not add any text, headers, or labels after the JSON.
2. You may wrap the JSON in a markdown code block (```json ... ```) or output raw JSON. Both are accepted.
3. Use strict JSON syntax: double quotes for all keys and string values; no trailing commas; no single quotes; no unescaped newlines inside strings.
4. All numeric fields must be integers (e.g. 85 not 85.0 or "85\%"). citation_precision_percentage and citation_completeness_percentage are numbers 0–100.
5. The "label" value must be exactly one of: "correct", "partially correct", "incorrect" (lowercase).
6. "missing_citations_details" must be a JSON array: use [] when there are none; each element must have "claim" and "should_cite" keys.
7. "justification" must be a single string. Escape internal double quotes and use \\n for newlines inside strings.
8. "verification.is_valid" must be boolean true or false, not a string.

Exact JSON schema (replace placeholders with your values):
{
  "summary": {
    "total_claims_identified": <integer>,
    "claims_with_citations": <integer>,
    "correctly_cited_direct": <integer>,
    "correctly_cited_derivable": <integer>,
    "incorrectly_cited": <integer>,
    "missing_citations": <integer>,
    "no_citation_needed": <integer>
  },
  "metrics": {
    "citation_precision_percentage": <integer 0-100>,
    "citation_completeness_percentage": <integer 0-100>
  },
  "verification": {
    "formula_check": "claims_with_citations = correctly_cited_direct + correctly_cited_derivable + incorrectly_cited",
    "is_valid": <true or false>
  },
  "missing_citations_details": [
    {"claim": "<claim text>", "should_cite": "<source(s) that should be cited>"}
  ],
  "label": "<correct|partially correct|incorrect>",
  "justification": "<detailed explanation covering: citation precision breakdown, missing citations and which sources should be cited, any incorrect citations, overall assessment and reasoning for label>"
}

REMINDER: Your response must end with this JSON block. Do not stop after the chain of thought. Without the JSON, the evaluation is invalid.

---

## Data

LLM Response:
<response>

Context:
<context>
\end{PromptBox}

\section{Hallucination Detection Task}
Model: Deepseek3.1

\label{app:hallucination}

\subsection*{System Prompt}

\begin{PromptBox}

You are a hallucination detection expert. Your role is to verify that information in LLM responses exists in the provided context.

Core Principles:

Verify EVERY claim against the context
Be LENIENT with formatting, rounding, and paraphrasing
Check for derivable values (conversions, approximations, calculations)
Only flag information that truly cannot be found or derived from context


CRITICAL: Your reply MUST end with the required JSON block (Part 2).  
Without it the evaluation cannot be recorded. Never omit the JSON.

\end{PromptBox}

\subsection*{User Prompt}
\begin{PromptBox}

Guiding Principle Ask: “Can I find the core information from the Response somewhere in the Context?”

ALL claims verified exactly in Context → factual (Score: 1.0)
MOST claims verified, or minor rounding discrepancies → partially factual (Score: 0.5)

MORE claims unverified than verified → hallucinated (Score: 0.0)
Verification Process

Step 1: Extract ALL Claims from Response
    Identify every factual claim:
    Numbers (prices, percentages, metrics, TVL, volumes)
    Dates and time periods
    Names (people, companies, organizations)
    Factual statements and assertions
    Any quantitative or specific data
    Ignore citations, formatting, and stylistic elements.
    
Step 2: Verify Each Claim Against Context
    For each claim:
        VERIFIED (exact): Exact match in Context
        VERIFIED (derivable): Can be derived via:
        Mathematical derivations
        Unit conversions
        Approximations
        Format variations
        Reasonable interpretations
        FABRICATED: Cannot be found or derived
        
Step 3: Verify Names and Entities
    Normalize (lowercase, remove punctuation)
    Allow spacing, capitalization, and suffix variations
    If normalized match → VERIFIED
    If no match → FABRICATED
    
Step 4: Verify Factual Claims
    Check whether the core substance exists in Context
    Paraphrasing and rewording are acceptable
    No contextual basis → FABRICATED
    
Step 5: Check for Contradictions
    If Response contradicts Context → CONTRADICTION (supports hallucinated label).
    
Step 6: Determine Final Label
    factual: All claims verified
    partially factual: Most claims verified
    hallucinated: More unverified than verified
    What Counts as Verified
    Exact matches, rounding, conversions, approximations
    Name variations and spacing differences
    Paraphrased but equivalent meaning
    Response Format
Part 1: Chain of Thought
CHAIN OF THOUGHT:

1. [Claim Name/Type]
   - Response claims: "..."
   - Context check: "..."
   - Derivation check: "..."
   - Status: VERIFIED / FABRICATED
Part 2: Structured Results (JSON) — Mandatory
{
  "summary": {
    "total_claims_identified": <integer>,
    "verified_exact": <integer>,
    "verified_derivable": <integer>,
    "fabricated": <integer>
  },
  "metrics": {
    "verification_rate_percentage": <integer>
  },
  "label": "factual | partially factual | hallucinated",
  "justification": "..."
}
Reminder: The response must end with the JSON block.
Without it, the evaluation is invalid.
\end{PromptBox}

\end{document}